\documentclass{ws-ijmpcs}

\usepackage{ifthen} 

\newboolean{uprightparticles}
\setboolean{uprightparticles}{false} 

\usepackage{lineno}  

\usepackage{xspace} 
\usepackage{upgreek}

\def\lhcb {\mbox{LHCb}\xspace}

\def\lhc    {\mbox{LHC}\xspace}

\def\MagUp {\mbox{\em Mag\kern -0.05em Up}\xspace}

\ifthenelse{\boolean{uprightparticles}}%
{

 \def\Pmu         {\ensuremath{\upmu}\xspace}

 \def\Ppi         {\ensuremath{\uppi}\xspace}

 \def\PDelta      {\ensuremath{\Delta}\xspace}                 
 \def\PXi      {\ensuremath{\Xi}\xspace}                 
 \def\PLambda      {\ensuremath{\Lambda}\xspace}                 
 \def\PSigma      {\ensuremath{\Sigma}\xspace}                 
 \def\POmega      {\ensuremath{\Omega}\xspace}                 
 \def\PUpsilon      {\ensuremath{\Upsilon}\xspace}                 
 
 \def\PB      {\ensuremath{\mathrm{B}}\xspace}                 
                  
 \def\PD      {\ensuremath{\mathrm{D}}\xspace}

 \def\PK      {\ensuremath{\mathrm{K}}\xspace}

 \def\Pi      {\ensuremath{\mathrm{i}}\xspace}

 \def\Ps      {\ensuremath{\mathrm{s}}\xspace}

}
{

 \def\Pmu         {\ensuremath{\mu}\xspace}

 \def\Ppi         {\ensuremath{\pi}\xspace}

 \mathchardef\PDelta="7101
 \mathchardef\PXi="7104
 \mathchardef\PLambda="7103
 \mathchardef\PSigma="7106
 \mathchardef\POmega="710A
 \mathchardef\PUpsilon="7107
                  
 \def\PB      {\ensuremath{B}\xspace}                 
                  
 \def\PD      {\ensuremath{D}\xspace}

 \def\PK      {\ensuremath{K}\xspace}

 \def\Pi      {\ensuremath{i}\xspace}

 \def\Ps      {\ensuremath{s}\xspace}

}

\DeclareRobustCommand{\optbar}[1]{\shortstack{{\miniscule (\rule[.5ex]{1.25em}{.18mm})}
  \\ [-.7ex] $#1$}}

\def\mun        {{\ensuremath{\Pmu^-}}\xspace} 

\def\squark    {{\ensuremath{\Ps}}\xspace}

\def\pion   {{\ensuremath{\Ppi}}\xspace}

\def\pip    {{\ensuremath{\pion^+}}\xspace}
\def\pim    {{\ensuremath{\pion^-}}\xspace}
\def\pipm   {{\ensuremath{\pion^\pm}}\xspace}

\def\kaon    {{\ensuremath{\PK}}\xspace}
  \def\Kbar    {{\kern 0.2em\overline{\kern -0.2em \PK}{}}\xspace}

\def\KorKbar    {\kern 0.18em\optbar{\kern -0.18em K}{}\xspace}

\def\Kp      {{\ensuremath{\kaon^+}}\xspace}
\def\Km      {{\ensuremath{\kaon^-}}\xspace}

\def\KS      {{\ensuremath{\kaon^0_{\mathrm{ \scriptscriptstyle S}}}}\xspace}

  \def\Dbar    {{\kern 0.2em\overline{\kern -0.2em \PD}{}}\xspace}
\def\D       {{\ensuremath{\PD}}\xspace}

\def\DorDbar    {\kern 0.18em\optbar{\kern -0.18em D}{}\xspace}
\def\Dz      {{\ensuremath{\D^0}}\xspace}
\def\Dzb     {{\ensuremath{\Dbar{}^0}}\xspace}

\def\Dpm     {{\ensuremath{\D^\pm}}\xspace}

\def\Dstarp  {{\ensuremath{\D^{*+}}}\xspace}

\def\Dspm    {{\ensuremath{\D^{\pm}_\squark}}\xspace}

\def\Bbar    {{\ensuremath{\kern 0.18em\overline{\kern -0.18em \PB}{}}}\xspace}

\def\BorBbar    {\kern 0.18em\optbar{\kern -0.18em B}{}\xspace}

  \def\Y#1S{\ensuremath{\PUpsilon{(#1S)}}\xspace}

\def\Lbar        {{\ensuremath{\kern 0.1em\overline{\kern -0.1em\PLambda}}}\xspace}
\def\LorLbar    {\kern 0.18em\optbar{\kern -0.18em \PLambda}{}\xspace}

\newcommand{\decay}[2]{\ensuremath{#1\!\to #2}\xspace}         

\def\to                 {\ensuremath{\rightarrow}\xspace}

\def\CP                {{\ensuremath{C\!P}}\xspace}

\newcommand{\ACP}{{\ensuremath{{\mathcal{A}}^{\CP}}}\xspace}

\def\AT#1     {\ensuremath{A_{\mathrm{T}}^{#1}}\xspace}           

\def\C#1      {\ensuremath{\mathcal{C}_{#1}}\xspace}                       
\def\Cp#1     {\ensuremath{\mathcal{C}_{#1}^{'}}\xspace}                    
\def\Ceff#1   {\ensuremath{\mathcal{C}_{#1}^{\mathrm{(eff)}}}\xspace}        
\def\Cpeff#1  {\ensuremath{\mathcal{C}_{#1}^{'\mathrm{(eff)}}}\xspace}       
\def\Ope#1    {\ensuremath{\mathcal{O}_{#1}}\xspace}                       
\def\Opep#1   {\ensuremath{\mathcal{O}_{#1}^{'}}\xspace}                    

\def\agamma     {\ensuremath{A_{\Gamma}}\xspace}

\newcommand{\tev}{\ifthenelse{\boolean{inbibliography}}{\ensuremath{~T\kern -0.05em eV}}{\ensuremath{\mathrm{\,Te\kern -0.1em V}}}\xspace}
\newcommand{\gev}{\ensuremath{\mathrm{\,Ge\kern -0.1em V}}\xspace}
\newcommand{\mev}{\ensuremath{\mathrm{\,Me\kern -0.1em V}}\xspace}
\newcommand{\kev}{\ensuremath{\mathrm{\,ke\kern -0.1em V}}\xspace}
\newcommand{\ev}{\ensuremath{\mathrm{\,e\kern -0.1em V}}\xspace}
\newcommand{\gevc}{\ensuremath{{\mathrm{\,Ge\kern -0.1em V\!/}c}}\xspace}
\newcommand{\mevc}{\ensuremath{{\mathrm{\,Me\kern -0.1em V\!/}c}}\xspace}
\newcommand{\gevcc}{\ensuremath{{\mathrm{\,Ge\kern -0.1em V\!/}c^2}}\xspace}
\newcommand{\gevgevcccc}{\ensuremath{{\mathrm{\,Ge\kern -0.1em V^2\!/}c^4}}\xspace}
\newcommand{\mevcc}{\ensuremath{{\mathrm{\,Me\kern -0.1em V\!/}c^2}}\xspace}

\def\mub{\ensuremath{{\mathrm{ \,\upmu b}}}\xspace}

\def\invfb   {\ensuremath{\mbox{\,fb}^{-1}}\xspace}

\def\deriv {\ensuremath{\mathrm{d}}}

\def\gsim{{~\raise.15em\hbox{$>$}\kern-.85em
          \lower.35em\hbox{$\sim$}~}\xspace}
\def\lsim{{~\raise.15em\hbox{$<$}\kern-.85em
          \lower.35em\hbox{$\sim$}~}\xspace}

\def\tell1  {TELL1\xspace}
\def\ukl1   {UKL1\xspace}

\newcommand{\ie}{\mbox{\itshape i.e.}\xspace}

\newcommand{\pis}{\ensuremath{\pi_{\rm s}}\xspace}
 
\newcommand{\DToEtapPi}{\ensuremath{\Dpm\rightarrow\Etap\pipm}} 
\newcommand{\DsToEtapPi}{\ensuremath{\Dspm\rightarrow\Etap\pipm}} 

\newcommand{\DToKsPi}{\ensuremath{\Dpm\rightarrow\KS\pipm}}

\newcommand{\DsToPhiPi}{\ensuremath{\Dspm\rightarrow\phi\pipm }}

\newcommand{\Etap}{\ensuremath{\eta^\prime}}

\usepackage{cite} 
\begin{document}

\markboth{Wojciech Krzemien}
{Mixing and CPV in charm hadrons at LHCb}

%
\catchline{}{}{}{}{}
%

\title{Mixing and CPV in charm hadrons at LHCb}

\author{Wojciech Krzemien\footnote{On behalf of the LHCb collaboration.}}
\address{High Energy Physics Division, National Centre for Nuclear Research, \\
05-400 Otwock-\'Swierk, Poland\\
wojciech.krzemien@ncbj.gov.pl}

\maketitle

\begin{history}
\published{Day Month Year}
\end{history}

\begin{abstract}
LHCb continues to expand its world-leading sample of charmed hadrons collected during LHC’s Run 1 (2010-2012) and Run 2 (2015-present). 
This sample is yielding some of the most stringent tests of the Standard Model understanding of charm physics.
This includes precise measurements of the neutral D-meson mixing parameters and some of the most sensitive searches for direct and indirect CP violation in charm interactions.
\end{abstract}

\section{Introduction}

Studies of charmed hadrons play an important role in the understanding of the strong and weak interactions
in the context of Standard Model (SM) and beyond.  
Precise measurements of the neutral \D-meson mixing parameters and searches for direct and indirect Charge-Parity violation (CPV) in charm interactions, 
are complementary to the analogous studies performed in the beauty and strange sectors.
The neutral D mesons are the only ones containing up-type quarks, which allows to probe a different quark dynamics with respect to other
neutral meson families.      
In the SM flavor mixing and the CP violation mechanisms are incorporated in the 3x3 unitary Cabibbo-Kobayashi-Maskawa (CKM) matrix, which parametrizes the transitions
between the three quarks families. The charm transitions at the tree level are described by the almost unitary and real 2x2 Cabbibo sub-space. 
Flavor mixing (and CP violation) may occur via loop processes that involves a flavor-changing-neutral-current (FCNC).
Indeed, the slow flavor mixing in D systems was confirmed experimentally in recent years. However no CP violation has been observed so far.
This result seems to confirm the SM predictions, in which CP violation effects are expected to be rather tiny (asymmetries not larger than $10^{-3}$). However,  New Physics (NP) contributions via loops 
could considerably enhance those effects. 
Since the SM processes are suppressed, the search for CP violation in the charm sector should provide good sensitivity for NP effects.
However, due to the relatively small mass of the charm quark, the theoretical calculations must cope with the long-distance, non-perturbative hadronic components, which makes the 
predictions rather imprecise.
To overcome these difficulties, high-statistics, precise experimental measurements in many decay modes are necessary. 

The \lhcb experiment operating at the Large Hadron Collider is well suited for such studies due to its excellent vertex resolution, excellent pion/kaon identification and 
its trigger scheme which allows to take advantage of the copious charm production in the proton-proton collisions at LHC ($\sigma(pp \rightarrow cc)\approx 1419 \mub$ at $\sqrt(s)=$7 TeV~\cite{sigma_7}
$\sigma(pp \rightarrow cc)\approx 2940 \mub$ at $\sqrt(s)=$ 13 TeV\cite{sigma_13}),
by the efficient selection of the high-purity hadronic state samples, while keeping the hadronic background at a low level~\cite{LHCb-detector}, \cite{LHCb-DP-2014-002},\cite{LHCb-DP-2014-004}.

The \lhcb experiment has already collected a world-leading sample of charmed hadrons collected during LHC’s Run 1 (2010-2012) and Run 2 (2015-present) and published various measurements of mixing parameters and CP violation observables
using $\Dz$ decays. 
In the next chapters the selected recent results on precise measurements of the neutral \D-meson mixing parameters and the sensitive direct and indirect CP violation searches will be briefly described.   

\section{Mixing and CP studies in decay}

The flavor eigenstates of the neutral charm mesons are not the (mass) eigenstates of the Hamiltonian that describes their time evolution. This fact 
leads to the oscillation phenomena (also called mixing) between the  $\Dz$ and $\Dzb$ mesons. CP violation in the linear superposition of flavor eigenstates defining the mass
eigenstates can lead to different mixing rates for $\Dz$ into $\Dzb$ and $\Dzb$ into $\Dz$.
The determination of mixing parameters and CP violation tests can be both performed by comparing the decay-time-dependent ratio of 
$\Dz\to K^+\pi^-$ to $\Dz\to K^-\pi^+$ rates with the corresponding ratio for the charge-conjugate processes.

The neutral \D-meson flavor at production is determined from the charge of the low-momentum pion (soft pion), $\pis^{+}$, produced in the flavor-conserving strong-interaction decay $\Dstarp\to\Dz\pis^+$.
 The right-sign (RS)  $\Dstarp\to\Dz(\to K^-\pi^+)\pis^+$ process is dominated by a Cabibbo-favored amplitude. Wrong-sign (WS) decays, $\Dstarp\to\Dz(\to K^+\pi^-)\pis^+$, arise from the doubly Cabibbo-suppressed $\Dz\to K^+\pi^-$ decay and the Cabibbo-favored $\Dzb\to K^+\pi^-$ decay that follows \Dz--\Dzb oscillation. Since the mixing parameters are small, $|x|,|y|\ll1$, the \CP-averaged decay-time-dependent ratio of WS-to-RS rates can be approximated as 
\begin{equation}\label{eq:true-ratio}
R(t) \approx R_D+\sqrt{R_D}\ y'\ \frac{t}{\tau}+\frac{x'^2+y'^2}{4}\left(\frac{t}{\tau}\right)^2,
\end{equation}
where $t$ is the proper decay time, $\tau$ is the average \Dz lifetime, and $R_D$ is the ratio of suppressed-to-favored decay rates. The parameters $x'$ and $y'$ are rotated by 
the strong-phase difference $\delta$ between the suppressed and favored amplitudes.
The equation is composed of (1) WS to RS ratio of non-mixed $\Dz$ mesons, (2) interference term and (3) mixing term, respectively.  
After the data selection, the efficiency-corrected WS-to-RS time-dependent experimental yield ratios are fitted under three hypothesis: no CP violation, no direct CP violation or all CP violation allowed. 
In the past years, the LHCb collaboration published several results exploiting this method, which among others, led to the first observation of the charm oscillation from a single measurement~\cite{LHCb-PAPER-2013-X}, excluding
the non-mixing hypothesis at the level of almost 10 $\sigma$.  

In the recent article, the results from the Run 1 ``prompt'' data sample where the \Dstarp meson is produced directly in $pp$
collisions~\cite{LHCb-PAPER-2013-053}, were combined with a Run 1 ``double-tagged''(DT) analysis, based on mesons produced in $ \overline{B} \to D^{*+}\mun X$, 
$D^{*+} \to D^0 \pi^+_s$, $D^0 \to K^{\pm} \pi^{\mp} $ process~\cite{LHCb-PAPER-2016-033}.
In this case, the flavor of the $\Dz$ at production is tagged twice, once by
the charge of the muon and once by the 
opposite charge of the slow
pion  $ \pi_s^+ $ produced in the $ D^{*+} $ decay, leading to very pure
samples.
Although the number of DT events constitutes only 2.5 \% of the full LHCb Run 1 statistics, 
simultaneously fitting the disjoint data sets of the two analyses
improves the precision of the measured parameters by 10\%--20\%. 
The combined fit of the DT and prompt sample is consistent with
\CP symmetry conservation.

The latest results in this channel from \lhcb include data from 2011-2016, corresponding to $5.0\invfb$, and an improved analysis method~\cite{LHCb-PRC-RC}. 
This update provides the most stringent
bounds on the parameters $A_D$ and $|q/p|$ from a single measurement, $A_D =(-0.1\pm9.1)\times10^{-3}$ and $1.00< |q/p| <1.35$ at the $68.3\%$ confidence level. 
Assuming \CP conservation, the mixing parameters are measured to be $x'^2=(3.9 \pm 2.7) \times10^{-5}$, $y'= (5.28 \pm 0.52) \times 10^{-3}$, and $R_D = (3.454 \pm 0.031)\times10^{-3}$. 
No evidence for \CP violation in charm mixing has been observed.

\section{Search for direct CPV with $D^{\pm}\rightarrow \eta' \pi^{\pm}$ and $D^{\pm}_{s}\rightarrow \eta' \pi^{\pm}$}

Another search  for  
for \CP\ violation, never measured before at the hadron collider,  has been recently performed~\cite{LHCb-PAPER-2017-Y} using 
$D^{\pm}\rightarrow \eta' \pi^{\pm}$ and 
$D^{\pm}_{s}\rightarrow \eta' \pi^{\pm}$ decays from 
proton-proton collision data. 
The overall experimental sample contained about 63000 $D^{\pm}$ and 152000 $D^{\pm}_{s}$ candidates,
which  corresponds to 
an integrated luminosity of 3\invfb,
recorded by the \lhcb experiment at centre-of-mass energies of $7$~and~$8\,$ TeV. 
The $\eta'$ meson is reconstructed via the $\pip \pim \gamma$ final state.
The CP-violation asymmetries were measured with respect to the control channels \DToKsPi\ and \DsToPhiPi\ to eliminate the detector
and production asymmetries. 

The obtained values of charge asymmetries are
$\ACP(\DToEtapPi)=(-0.61\pm 0.72 \pm 0.53 \pm 0.12)\%$~and $\ACP(\DsToEtapPi)=(-0.82\pm 0.36 \pm 0.22 \pm 0.27)\%$, where 
the first uncertainties are statistical, the second systematic, and the third 
are the uncertainties on the 
$\ACP(\DToKsPi)$ and $\ACP(\DsToPhiPi)$ measurements used for calibration. 
The results represent the most precise measurements of these 
asymmetries to date, and are  
consistent with CP symmetry invariance.

\section{Indirect CP violation searches in \decay{\Dz}{h^{+}h^{-}} measurements}


Determination of indirect asymmetry of effective lifetimes~\cite{LHCb-PAPER-2017-X} in the singly-Cabibbo-suppressed
decays into CP-eigenstates, $f$, where $f=\pi^+\pi^-$ or $f=K^+K^-$, has been performed by
a new, independent analysis method with respect to the previously published result~\cite{LHCb-PAPER-2013-054}.

The time-dependent \CP asymmetry of the studied modes can be expressed as~\cite{Aaltonen:2011se}:  
\begin{equation}
A_{\CP}(t) \equiv \frac{\Gamma(\decay{\Dz(t)}{ f}) - %
 	 			\Gamma(\decay{\Dzb(t)}{ f})}%
				{\Gamma(\decay{\Dz(t)}{ f}) + %
 				\Gamma(\decay{\Dzb(t)}{ f})} 
		\simeq a^f_\textup{dir} -\agamma \frac{t}{\tau_D}, 
\label{eq:Aind}
\end{equation}
where $\Gamma (\Dz(t) \to f)$ and $\Gamma (\Dzb(t) \to f)$ indicate
the time-dependent decay rates of an initial \Dz or \Dzb decaying to a final state $f$ at decay time $t$,
$\tau_D = 1/\Gamma = 2/(\Gamma_1 + \Gamma_2)$ is the average lifetime of the \Dz meson, 
$a^{f}_\textup{dir}$ is the asymmetry related to direct \CP violation and
$\agamma$ is the asymmetry between the \Dz and \Dzb effective decay widths,
\begin{equation}\label{eq:A_G}
\agamma \equiv \frac{\hat{\Gamma}_{\decay{\Dz}{ f}} - %
 	 			\hat{\Gamma}_{\decay{\Dzb}{ f}}}%
				{\hat{\Gamma}_{\decay{\Dz}{ f}} + %
 				\hat{\Gamma}_{\decay{\Dzb}{ f}}}. 
\end{equation}
The effective decay width $\hat{\Gamma}_{\decay{\Dz}{ f}}$ 
is defined as $  \int_0^\infty \Gamma( \Dz(t) \to f) \, \deriv t  /
\int_0^\infty t \, \Gamma(\Dz(t) \to f)\, \deriv t$, \ie the inverse of
the effective lifetime.
Neglecting contributions from subleading 
amplitudes ,
$a^{f}_\textup{dir}$ vanishes and \agamma is independent of the final state $f$.

Asymmetries in the time-dependent rates of \decay{\Dz}{\Kp \Km} and \decay{\Dz}{\pip \pim} 
decays are measured in data sample collected with the \lhcb detector during \lhc Run 1.
The reconstructed events data distributions are
corrected for detector non-uniformities and the presence of secondary decays, and the raw asymmetry is
calculated in bins of $\Dz$ decay time. Next the yields  are extracted, and 
a linear fit of the A raw asymmetry distribution as a function of decay time is used to extract 
$\agamma$.

The asymmetries in effective decay widths between 
\Dz and \Dzb decays are
measured to be
$\agamma(\Kp \Km) = (-0.30 \pm 0.32 \pm 0.10)\times 10^{-3}$ and 
$\agamma(\pip \pim) = (0.46 \pm 0.58 \pm 0.12)\times 10^{-3}$, 
where the first uncertainty is statistical and the second systematic.
The results for \decay{\Dz}{\Kp\Km} and \decay{\Dz}{\pip\pim} are consistent 
and show no evidence of \CP violation. 
Assuming that only indirect \CP violation contributes to \agamma,
and accounting for correlations between the systematic uncertainties,   
the two values 
can be averaged to yield a single value of
$\agamma = (-0.13 \pm 0.28 \pm 0.10)\times 10^{-3}$, 
while their difference is 
$\Delta \agamma = (-0.76 \pm 0.66 \pm 0.04)\times 10^{-3}$.
The above average is consistent with the result obtained by \lhcb in a muon-tagged sample~\cite{LHCb-PAPER-2014-069}, 
which is statistically independent. 
The two results are therefore combined to yield an 
overall \lhcb Run 1 value 
$\agamma = (-0.29 \pm  0.28)\times 10^{-3}$ 
for the average of the $\Kp\Km$ and  $\pip\pim$ modes.

\begin{figure}[pb]
\centerline{\includegraphics[width=9cm]{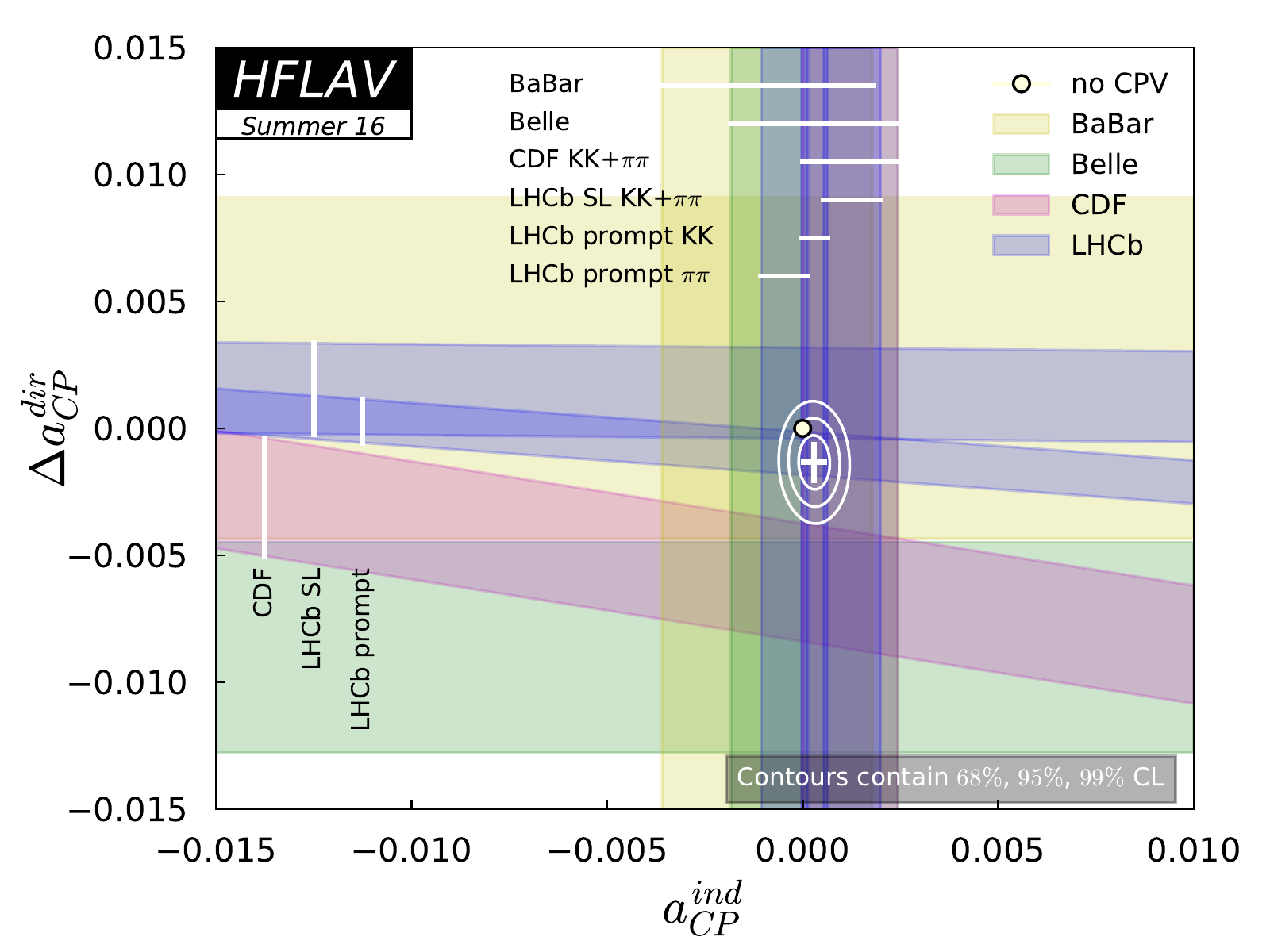}}
\vspace*{8pt}
\caption{The indirect vs direct CP parameters from independent experimental measurements, together with the world average denoted as white cross. The white dot corresponds to the no CPV hypothesis.  The Figure is adopted from~\cite{HFLAV}. \label{f1}}
\end{figure}

The presented result can be confronted with other experimental measurements by looking at the combination of all independent CP searches, taken from ~\cite{HFLAV} and presented in the Fig.~\ref{f1}.
The global fit to all data is consistent with the no-CP violation hypothesis.

\section{Summary}

The measurement of the mixing parameters and the searches of CPV in the charm sector provide
a precise tests of the SM and probes of New Physics effects, complementary to analogous searches 
with B and K mesons. 

The \lhcb experiment has used a world-leading sample of charm particles to provide a set of measurements 
based
on the Run 1 data which confirmed the SM predictions. 
The no-mixing hypothesis for $\Dz$ system has been excluded at more than 10 $\sigma$ level,  based on data gathered
in Run 1. Results based on both prompt and double-tagged candidates are in agreement.
All measurements are consistent with the no-CP violation hypothesis. The current precision reached is of
order of $10^{-4}$ for the indirect searches.  

The majority of the measurements are statistically limited, and some of the systematics factors are expected to 
reduce with signal yields. Currently, several Run~1 and Run~2 based analyses are ongoing. 
The  Run~2 measurements benefits not only from the higher statistics but also from the optimized trigger.

\section{Acknowledgements}
I express my gratitude to the National Science Centre (NCN) in Poland
for the financial support under the contract 2014/12/S/ST2/00459.

\end{document}